\documentstyle[12pt]{article}
\begin{document}

\begin{titlepage} 

\begin{flushright}
CERN-TH/2000-375 \\
hep-th/0012216 \\
\end{flushright}

\vspace{0.5in}

\begin{centering} 

{\large \bf Space-Time Foam Effects on Particle Interactions and the GZK
Cutoff}

\vspace{0.5in}

{\bf John~Ellis}$^a$, {\bf N.E.~Mavromatos}$^{a,b}$ and  
{\bf D.V.~Nanopoulos}$^{c}$

\vspace{0.5in} 
{\bf Abstract}

\end{centering} 

{\small 

Modelling space-time foam using a non-critical Liouville-string model for
the quantum fluctuations of $D$-branes with recoil, we discuss the issues
of momentum and energy conservation in particle propagation and
interactions. We argue that momentum should be conserved {\it exactly}
during propagation and {\it on the average} during interactions, but that
energy is conserved only {\it on the average} during propagation and {\it
is in general not} conserved during particle interactions, because of
changes in the background metric.  We discuss the possible modification of
the GZK cutoff on high-energy cosmic rays, in the light of this energy
non-conservation as well as the possible modification of the usual
relativistic momentum-energy relation. 

}

\vspace{0.2in}
\begin{flushleft} 

$^{a}$ CERN, Theory Division, CH 1211 Geneva 23, Switzerland. \\
{~~}\\
$^{b}$ Department of Physics, King's College London, 
Strand, London WC2R 2LS, United Kingdom.\\
{~~}\\
$^{c}$ Department of Physics, Texas A \& M University, 
College Station, TX~77843, USA; \\
Astroparticle Physics Group, Houston
Advanced Research Center (HARC), 
Mitchell Campus,
Woodlands, TX~77381, USA; \\
Chair of Theoretical Physics, 
Academy of Athens, 
Division of Natural Sciences, 
28~Panepistimiou Avenue, 
Athens 10679, Greece. \\
\end{flushleft} 

\end{titlepage}

\newpage

\section{Introduction}

Analysts of quantum gravity expect that the fabric of space-time should
fluctuate over distance scales of the order of the Planck length $\ell_P$
and over time scales of the order of the Planck time $t_P$~\cite{foam}.  
The Big Issue
is whether there might be any observable consequences of this `space-time
foam'.  Various ways of modelling certain aspects of quantum fluctuations
in space-time have been proposed, including loop gravity~\cite{loop}, 
non-critical
Liouville strings~\cite{ddk,emn} and $D$-branes~\cite{dbranes}.  
Within the latter framework, it has
been suggested~\cite{emn} that entropy 
might not be conserved at the microscopic
level, during both particle propagation and interactions, and that energy
might be conserved at best statistically. It was also on the basis of this
calculational scheme that we proposed that the conventional
special-relativistic relation between momentum and energy might be
modified, reflecting a breakdown of Lorentz invariance and leading to an
effective refractive index {\it in vacuo}~\cite{aemn,nature}. 
Related effects on particle
propagation, such as birefringence, have been suggested in the context of
the loop approach to quantum gravity~\cite{pullin}.

Subsequently and independently of these theoretical developments, it has
been pointed out
that a modification of the conventional special-relativistic relation
between momentum and energy might dissolve the Greisen-Zatsepin-Kuzmin
(GZK)
cutoff~\cite{GZK} on high-energy cosmic rays~\cite{mestres,grillo}.  
Simply put, if $E \ne p$ for
ultra-relativistic particles, the kinematical thresholds for reactions
such as $N + \gamma \rightarrow \Delta \rightarrow N + \pi$ and $\gamma +
\gamma \rightarrow e^- + e^+$ would be modified, altering the absorption
lengths of ultra-high-energy nucleons $N$ and photons $\gamma$. The two
lines of thought were brought together by Kifune~\cite{kifune}, 
who pointed out that the
modification of the energy-momentum relation proposed in the Liouville
$D$-brane approach~\cite{emn,aemn} might remove 
the GZK cutoff for ultra-high-energy
photons striking the cosmic microwave background radiation, and by
Protheroe and Meyer~\cite{protheroe}, 
who made a similar observation for TeV photons
striking the astrophysical infra-red background. 

Previous authors have not considered the possibility of energy and/or
momentum non-conservation during particle {\it interactions}, such as $N +
\gamma \rightarrow \Delta \rightarrow N + \pi$ or $\gamma + \gamma
\rightarrow e^- + e^+$, which could also affect their kinematical
thresholds, and hence the corresponding absorption lengths and GZK
cutoffs. In this paper, we clarify the circumstances under which we would
expect, on the basis of our Liouville-string $D$-brane model for
space-time foam, whether and when the energy and/or momentum of observable
particles should not be conserved, and by what amount. We extend our
previous analyses to include particle interactions, and comment on the
implications for analyses of the GZK cutoffs.

We first review our arguments that, during propagation, the momentum of an
observable particle should be conserved {\it exactly} and its energy
conserved {\it statistically}. On the other hand, we argue in this paper
that, during interactions, the sum of the momenta of observable particles
should only be conserved {\it statistically}, and their total energy {\it
is not in general} conserved. The basic intuition behind the conservation
of energy and momentum during particle propagation is that, although the
energy of a particle modifies the background metric as `recoil' effect,
and this metric modification in turn modifies the relativistic
momentum-energy dispersion relation, this modification is space- and
time-translation invariant. The particle feels an `unusual' non-flat
metric, but this is a constant of motion. On the contrary, the `recoil'
background metric changes during an interaction, leading in general to a
violation of energy conservation. However, momentum remains conserved
statistically, as a consequence of basic properties of correlation
functions in Liouville string theory.  The lack of energy conservation
during particle interactions modifies the kinematical threshold for
particle production quantitatively, but the qualitative effects on
particle absorption lengths and the GZK cutoffs~\cite{kifune} still
remain.

\section{Particle Propagation: Modified Dispersion Relations and
Energy-Momentum Conservation} 

We review in this Section our use of the Liouville-string $D$-brane model
of~\cite{emn} to
discuss the
propagation of a closed-string state in the background of a
single $D$ particle, which we interpret as a model of (matter) particle
propagation through a (dilute) space-time foam. The recoil of the
$D$ particle during the scattering
distorts the surrounding space-time, inducing a non-trivial
off-diagonal term in the metric~\cite{emn}: 
\begin{equation}
G_{0i}=-u_i=-(p_{0,i}-p_i)/M_D~, i=1,\dots.  
\label{metric}
\end{equation} 
where $u_i$ denotes the recoil velocity of the $D$ particle, 
where $i=1,\dots, d(=3)$ is a spatial index, 
$M_D \sim 10^{19}$~GeV is the quantum-gravity scale,
and $p_0$ ($p$) denotes the spatial momentum of the string state before
(after) the scattering,
in the frame where the $D$ particle is initially at rest~\footnote{Throughout 
this work, we use units where the (low-energy) light velocity 
{\it in vacuo} is $c_0=1$.}.

After the collision, the string propagates in a background space-time
with metric (\ref{metric}). We consider first the case of 
massless particles.
The on-shell relation for a massless point-like
particle in the metric (\ref{metric}) reads:
$p_\mu p_\nu g^{\mu\nu}=0 \to -E^2 +p_i^2 + Ep_i u_i =0$,
from which we derive
\begin{equation}
E=|p_i|\left(1 + \frac{1}{4}(p_i u_i)^2\right)^{1/2} + \frac{1}{2}p_i u_i
\label{firstpart}
\end{equation}
Using $u_i=\frac{p_o-p_i}{M_D}$, we then find that
\begin{equation}
E = |p_i| + \frac{1}{2M_D}p_i(p_{0,i}-p_i) + {\cal O}(\frac{1}{M_D^2})
\label{dispersion} 
\end{equation} 
We consider an average $<< \dots >>$ over {\it both}: (i) statistical 
effects due to the sum over a gas of $D$ particles 
in the background, as is appropriate for a space-time foam picture,
and (ii) quantum effects, which are treated by summation over higher 
world-sheet topologies.
We assume a random-walk model, 
averaging over the
angle between the incident and scattered particles.
Randomness implies that 
$<<p_{0,i}p_i>> = 0$ and $<<u_i>> = 0$. 
Note that, 
although the metric perturbation $h_{\mu\nu}$ 
(\ref{metric}) vanishes on the average, 
its two-point correlation does not vanish:
\begin{equation} 
<<h_{\mu\nu}>>=0~, \qquad 
<<h_{\mu\nu}h_{\rho\sigma}>> \ne 0
\label{isotropic}
\end{equation} 
as recently found also in other stochastic gravity models~\cite{ford}. 

Because of momentum conservation during the scattering
of the closed string with the $D$ particle, which is rigorously
true within the world-sheet $\sigma$-model approach, we find that
\begin{equation}
<<p_0 - p>>_i = <<M_D u_i>>=0
\label{momentumconservation}
\end{equation}
i.e., spatial momentum is conserved on the average. It is easy to see
that this holds for an arbitrary number of local
interactions involving $N$ matter particles:
\begin{equation}
<< \sum_{n=1}^{N} p_i >>= M_D <<u_i>>=0,
\label{momconsn}
\end{equation}
which is an important property of our space-time foam model that we use
in the following. 

Let us now consider the variance
\begin{eqnarray}
(\Delta p_i)^2 = <<p_ip_i>> - <<p_i>>^2,
\label{variance}
\end{eqnarray}
where we recall that the summation in $<< \dots >>$ includes both quantum
and
statistical effects. We return later to 
a discussion of its possible value in a specific stringy model.

Performing the average $<<\dots >>$ of equation
(\ref{dispersion}),
we obtain~\cite{emn}:
\begin{equation} 
<<E>>\equiv {\overline E} = {\overline p}
\left(1 -\frac{1}{2M_D}{\overline p} + \dots \right)~:
~~{\overline p} \equiv <<|p_i|>>.
\label{averagedisp1}
\end{equation} 
To make it clear that the
statistical averaging, which depends on the details of the foam,
cannot be performed quantitatively in terms of any known $D$-brane mass
scale $M_D$, we
replace $1/2M_D$ by $\xi/2M_D$, where we expect $\xi ={\cal O}(1)$ for a
dilute-gas foam model.
Thus, the following modified dispersion relation characterizes our foamy
model:
\begin{equation}   
{\overline E} = {\overline p}
\left(1 -\frac{\xi}{2M_D}{\overline p} + \dots \right)~, 
~\xi >0.
\label{averagedisp}
\end{equation}
This corresponds to a non-trivial refractive index, which is
subluminal, as expected from the Born-Infeld dynamics of the $D$-brane
foam:
\begin{equation}
c({\overline E})=\frac{\partial {\overline E}}{\partial {\overline p}}
= 1 - \frac{\xi}{M_D}{\overline p}.
\label{refractive}
\end{equation}
We have discussed elsewhere~\cite{nature,mitsou} how such a phenomenon can
be tested using
Gamma-Ray Bursters and other intense astrophysical probes~\cite{others},
by looking
for delays in the arrival times of photons with different energies.

One may study in a similar manner the modification of the
dispersion relation
for a particle of mass $m \ne 0$. The on-shell relation now reads:
$p_\mu p_\nu g^{\mu\nu}= - m^2 \to -E^2 +p_i^2 + Ep_i u_i  + m^2 =0$,
from which we infer,
following an approach similar to the massless case in which we
assume isotropic random foam,
\begin{equation} 
{\overline E} = {\overline p} - \frac{\xi}{2M_D} {\overline p}^2 
+ <<\frac{m^2}{2 |p_i |}>> + \dots~, \qquad {\overline p}=<<|p_i|>> 
\label{massivedisp}
\end{equation} 
for highly-energetic massive particles.
The quantity $<<\frac{m^2}{|p_i|}>>$ is not trivial to evaluate, because
we have to invoke the precise meaning of a quantum-fluctuating
momentum in our $\sigma$-model approach. 
We know that, in the case of a $D$ particle, momentum can be
represented as a coupling
in the $\sigma$ model, whose tree-level value represents the
average momentum ${\overline p}$, and the summation over world-sheet topologies
can be shown to correspond, in leading order, to Gaussian fluctuations 
$\delta p_i$. Working to leading 
order in small flucutations, we find $<<m^2/|p_i|>>\simeq m^2/{\overline
p}$. A more detailed analysis
is pending, but this assumption will prove sufficient for our 
qualititative description, so we therefore write the following
modified dispersion relation
\begin{equation} 
{\overline E} = {\overline p} - \frac{\xi}{2M_D} {\overline p}^2 
+ \frac{m^2}{2{\overline p}} + \dots~, \qquad {\overline p}=<<|p_i|>> 
\label{massivedispersion}
\end{equation}
for a highly-energetic massive particle.

We now consider the issue of energy conservation during particle
propagation through space-time foam.
We consider a particle with incident (final) energy-momentum 
$(E_{1(2)},p_{1(2)})$ scattering off a $D$ particle in an isotropic foamy
model of the type introduced above.
For simplicity, we concentrate on one-dimensional propagation:
the extension to higher dimensions is straightforward. 
The quantity $\delta E=E_1-E_2$  measures energy non-conservation
in a single scatering. Averaging over both quantum fluctuations 
and statistical foam effects, and using (\ref{averagedisp}), one has:
\begin{eqnarray}
&~& <<\delta E>>={\overline E}_1 - {\overline E}_2 =
{\overline p}_1 - {\overline p}_2 - \nonumber \\
&~& \frac{\xi}{2M_D}\left((\Delta p_1)^2 -
(\Delta p_2)^2 + {\overline p}_1^2 - {\overline p}_2^2 \right) = 0,
\label{energyconser}
\end{eqnarray}
where we have taken into account (\ref{momentumconservation}) 
and we assume that
the momentum variances of the incoming and outgoing 
particles 
are of the same order, as one would expect for energies $E << M_D$.

Thus we arrive at the following conclusion: 

\noindent {\bf Theorem 1}:

{\it Momentum is conserved exactly and energy is conserved 
on the average in the propagation of observable
matter particles in isotropic models of Liouville $D$-particle
foam. However, as has been emphasized previously, this
energy conservation {\it is not absolute}: it is a statement about
expectation values, not an operator statement}.

As also emphasized previously, this statistical energy conservation is a
consequence of the renormalizability of the world-sheet $\sigma$-model
theory. The statistical variance in the energy measured for a photon of
given
momentum may be related to quantum (loop) effects in the $\sigma$-model
treatment of the $D$ brane recoil~\cite{emnstoch}. We do not discuss this
effect here in
any detail, but note the consclusion, namely that we expect a variation
$\Delta E$ in the measured energy of the generic form
\begin{equation}
\Delta E \, = \, g_s \frac{\zeta}{2 M_D} {\overline p}^2
\label{variation}
\end{equation}
where we have indicated explicitly a factor of the string coupling $g_s$,
and emphasize that the numerical coefficient $\zeta$ is unknown and
distinct from the corresponding parameter $\xi$ in the modified
dispersion relation (\ref{massivedispersion}). In a weakly-coupled
string model, the variation (\ref{variation}) might well be negligible,
but it might be important in a strongly-coupled model. If it is important,
it would have the effect of smearing out the effect on the GZK cutoff
that we discuss later.

\section{Particle Interactions in Liouville Space-Time Foam}

We now extend the previous analysis to consider the decay of a
highly-energetic 
particle with energy and momentum $(E_1,p_1)$ 
into two other massless particles with energy-momentum vectors
$(E_2,p_2)$ and $(E_3,p_3)$,
in the context of our Liouville-string approach. 
Particle interactions are represented
as correlators among vertex operators $V_i$ in the world-sheet $\sigma$
model,
which represent the various particle excitations. 
It is known~\cite{adler} that, in the context of 
non-critical Liouville strings~\cite{ddk},
an $N$-point correlator of such vertex operators,
${\cal F}_N \equiv <V_{i_1} \dots V_{i_N}>$, where $<\dots>$ 
signifies a world-sheet
expectation value in the standard Polyakov treatment, 
transforms 
under infinitesimal Weyl shifts of the world-sheet metric 
in the following way: 
\begin{equation}
\delta_{\rm weyl} {\cal F}_N \, = \, 
\left[ \delta_0
+ {\cal O}\left(\frac{s}{A}\right)\right]{\cal F}_N
\label{weyl}
\end{equation}
where the standard part $\delta_0$ of the variation
involves a sum over 
the conformal dimensions $h_i$ of the operators $V_i$ that is independent
on the world-sheet area $A$. The quantity $s$ is the sum of 
the gravitational anomalous dimensions~\cite{ddk}:
\begin{equation}
s=-\sum_{i=1}^{N}\frac{\alpha_i}{\alpha} - \frac{Q}{\alpha}~,
~\alpha_i=-\frac{Q}{2}+ \frac{1}{2}\sqrt{Q^2 + 4(h_i-2)},
~\alpha=-\frac{Q}{2}+ \frac{1}{2}\sqrt{Q^2 + 8}
\label{gad}
\end{equation}
where $Q$ is the central-charge deficit. 

In standard critical
string theory, the vertex operators corresponding to 
excitations are $\propto e^{ik_i X^i}$, where $k_i$ is the 
momentum and the index $i$ runs over both space and time.
However,
in the non-critical treatment~\cite{emn} in which time is
identified with the zero mode of the
Liouville field $\phi$, the index $i$ is {\it only spatial}. 
The role of the energy is played in this approach by the gravitational
anomalous dimension $\alpha_i$, since the dressed excitation 
vertex $V_i^L$ is~\footnote{We recall that Liouville dressing is
necessary 
in order to restore 
world-sheet conformal invariance (criticality)~\cite{ddk}.}:
\begin{equation}
V_i^L \sim e^{\alpha_i\phi} e^{ik_iX^i}
\label{dressed}
\end{equation} 
It is then evident that the
integration over the
$\sigma$-model spatial coordinates in the world-sheet correlator ${\cal
F}_N$, will always imply 
conservation of the {\it spatial momentum}. 

There is a caveat in
this argument, namely that the form (\ref{dressed}) of the vertex
operator describes an excitation on a flat space-time.  This geometry
is only an {\it average} effect in models of isotropic space-time
foam (\ref{isotropic}). In general, quantum-gravity effects caused, e.g., 
by the recoil of a heavy $D$ particle as we are considering here,
distort the space-time. This has the inevitable result that the
plane-wave expansion (\ref{dressed}) breaks down.
The corresponding conservation of momentum therefore is 
demoted to an average effect due to the statistical sum over space-time
foam fluctuations.

Turning now to the energy, we observe from (\ref{dressed}) that its
conservation is not immediately apparent, because
the (Liouville) energies
are weighted by a real exponent in the exponential
rather than a phase factor.
However~\cite{emn,adler}, 
there is an alternative way of imposing Liouville 
energy conservation in the mean, which follows from a
careful treatment of (\ref{weyl}), interpreting
the (world-sheet) zero mode of the Liouville field, $\phi_0$, as 
target time $t$. This zero mode appears inside the world-sheet
covariant area $A$~\cite{adler}:
$\phi_0 \propto A^{1/2}$. 

If $s \ne 0$, i.e., the Liouville energy is not conserved,
one obtains a non-trivial 
time dependence on the associated correlator ${\cal F}_N$,
which means that it can no longer be interpreted as a unitary
$S$-matrix amplitude in a factorizable $\$$ matrix element~\cite{emn,adler}. 
On the other hand,
the ordinary kinematics 
of particle interactions, that is commonly assumed in most approaches
to the phenomenology of quantum gravity~\cite{mestres,grillo,kifune,protheroe},
holds only in models in which the non-criticality of the string 
does not imply a breakdown of $\$$-matrix factorization
when the quantum-gravitational interactions are properly 
taken into account. This means that the complete matter + gravity
system should behave like an ordinary quantum-mechanical 
system, as in conventional critical string theory.

This is not always feasible in practice, 
and in generic Liouville theory a precise mathematical 
description of quantum-gravity induced effects 
is still lacking.
However, this is possible in
simplified models of $D$-brane foam~\cite{emn}, 
where
foamy effects are represented by looking at the back-reaction effects 
on the space-time geometry that arise from its 
distortion during the scattering of a propagating stringy matter
mode on a heavy, non-relativistic $D$ particle embedded 
in the space-time. In such a model, the 
recoil of the $D$-particle defect is described by a 
logarithmic conformal field theory on the world sheet. In this framework,
the closed world sheet of the propagating stringy matter particle is 
torn apart by the presence of the defect
during the scattering event~\cite{emn}, 
and the recoil of the defect is represented as a (non-conformal)
logarithmic deformation of the world sheet~\cite{kmw}.
Dressing with the Liouville field restores criticality, but also 
leads to a bulk gravitational field (\ref{metric}), which expresses the 
distortion of the space-time surrounding the scattered defect. 

As a result of the recoil treatment, the effective stringy
$\sigma$ model becomes non-critical, since the
corresponding deformations
are relevant from a world-sheet renormalization group 
viewpoint~\cite{kmw}. This is the exclusive source of non-criticality,
which implies that the induced central-charge deficit $Q$ 
can be expressed in terms of the corresponding deformation 
couplings. The leading efffect comes from the deformation associated
with the recoil velocity $u_i$ (\ref{metric}), 
and is given by the 
kinetic energy of the recoiling $D$ particle, to leading order in an
approximation of
small recoil velocity, as appropriate for non-relativistic heavy 
$D$ particles. To develop this approach, we make the following formal steps.

(i) We first consider the gravitational anomalous  dimension
$\alpha_D$ 
of a recoiling $D$-particle excitation. In Liouville theory~\cite{ddk},
there are two possibilities: 
\begin{equation} 
\alpha_D^{(\pm)} = -\frac{Q}{2} \pm \sqrt{\frac{Q^2}{4} + 
\frac{\epsilon^2}{2}}  
\label{ad}
\end{equation}
where we took into account the fact~\cite{kmw} that 
the deformation of a recoiling $D$ particle 
has an anomalous dimension $-\frac{\epsilon^2}{2}$
on a flat world-sheet, satisfying:  
\begin{equation} 
\epsilon \sim {\rm ln}(L/a)\to 0
\label{epsrel}
\end{equation} 
where $L (a)$ is the infrared
(ultraviolet) world-sheet distance cutoff, and the relation (\ref{epsrel})
between $\epsilon$ and the cutoff is dictated by the closure
of the logarithmic world-sheet algebra. 

Then (ii) in standard Liouville
theory, for excitations that appear
as external legs in string amplitudes, it is only the $\alpha^{(+)}$ 
excitations that are kept. However, the $D$-particle 
recoil excitations, since they express virtual foam effects, 
can only appear as internal lines of string amplitudes, 
so the $\alpha_D^{(-)}$ solutions are in principle {\it allowed}. 
In fact, it is the $\alpha _D^{(-)}$ choice that should be selected
in  order to obtain statistical energy conservation, as we discuss next.

(iii) To this end, we consider the sum of the anomalous
dimensions of $N$ matter particles plus a recoiling $D$ particle
excitation, with gravitational dimension $\alpha_D^{(-)}$,
which contributes to the central charge deficit $Q^2=C-C^*$,
where $C^*$ represents the critical-string central charge, e.g., $C^*=25$
in bosonic strings. The deficit $C$ 
is calculated at the end of the scattering with the $D$ particle, 
which, in the terminology of~\cite{kmw}, occurs at a (large)  time 
$t\sim 1/\epsilon^2$,
consistent with (\ref{epsrel}) and 
the identification of the Liouville mode with the 
target time~\cite{emn}.
The deficit $C$ is given~\cite{ms}  by
the standard Zamolodchikov $C$ theorem~\cite{zam} applied to 
the open sector of the $\sigma$ model~\footnote{We recall that the 
Liouville dressing is applied to the closed sector, by rewriting the
boundary recoil deformation as a 
total derivative on the bulk world sheet. In order to  
compute the central charge deficit $Q$, which is an entity 
that exists prior to Liouville dressing, to leading order one has to take
into account only the boundary conformal field theory of the recoil 
deformations, and apply 
Zamolodchikov's $C$-theorem there.}:
$C \simeq C^* + \int_{t_0}^t \beta^i {\cal G}_{ij} \beta^j $,
where the indices $i,j=C,D$ run over the pair of recoil operators
$V_i$ 
describing the collective coordinates (position and momenta) 
of the recoiling $D$ particle~\cite{kmw}, both with anomalous dimensions
$-\epsilon^2/2$ in flat world sheets.
The function ${\cal G}_{ij} \sim <V_i V_j>$ is the 
Zamolodchikov metric in theory space of these operators~\cite{zam},
whilst the 
world-sheet renormalization-group $\beta$ functions are
written as: $\beta^i = -\frac{\epsilon^2}{2}g^i + {\hat \beta}^i$,
and ${\cal G}_{ij} \sim \frac{1}{\epsilon^2}{\cal G}_{ij}^{(1)} + \dots$,
as in standard theory~\cite{emn}, with ${\cal G}_{ij}^{(1)}$
the residue of the simple poles in $\epsilon^2$. 
In the problem at hand, one has ${\hat \beta}^C=u^i $, ${\hat \beta}^D=0$,
${\cal G}_{CC}^{(1)}\propto -\epsilon^2$ and ${\cal G}_{CD}^{(1)} \sim 1$.  
Hence, to leading order in $\epsilon^2 \to 0$, the 
central charge deficit $C$ is then given by 
\begin{equation}
Q^2 = Q_0^2 + \frac{1}{\epsilon^2}(u^i)^2~, , ~\qquad |u_i| << 1   
\label{ccd}
\end{equation} 
where $Q_0^2$ is a combination of constants 
which identifies the `critical' theory, defined 
as the one with $u_i\to 0$, i.e., no recoil. 
In the $u\to 0$ limit, the central charge of the 
$\sigma$ model may be identified 
with that of a four-dimensional string theory, and the deficit 
$Q$ is then just a numerical constant, corresponding to a 
non-trivial dilaton background that varies linearly 
with the Liouville zero mode/time.
Such theories with linear dilatons have been considered 
in~\cite{aben}, where it was argued that they provide examples of 
expanding universes in string theory. However,
simple linear dilaton models with flat metric space-times, 
such as the ones derived 
in the limit $u\to 0$ above, actually describe a non-expanding 
string universe if the lengths are measured by string rods~\cite{sanchez}.  
In this sense, the critical four-dimensional model obtained in 
the no-recoil limit $u \to 0$ describes an `equilibrium' situation. 

Finally, (iv) the constant $Q_0^2$ on the r.h.s. of (\ref{ccd})
will be {\it fixed} by the requirement
of {\it energy conservation} in the complete matter + 
recoiling $D$-brane system, as we discuss below. 

\section{Energy Non-Conservation during Interactions} 

We now examine the question whether energy is conserved during
interactions, in the $D$-particle foam picture outlined above.
Consider $s=0$ in (\ref{gad}), where in $s$ one now includes the 
$D$-particle anomalous dimension $\alpha_D^{(-)}$, with $Q$ given by
(\ref{ccd}). According to the
previous discussion, this guarantees unitarity of the complete 
matter + recoiling $D$-particle system.
Expanding the appropriate denominators up to leading order in 
$(u^i)^2/\epsilon^2$, one obtains 
\begin{equation} 
\sum_{i=1}^{N} E_i \simeq \frac{1}{2}M_D u^2 
\label{liouvener}
\end{equation}
where $i=1,\dots, N$ runs over all the particles in the
interaction, e.g.,
$N=3$ in the $1 \to 2$-body decay process considered here,
and the $E_i$ denote the corresponding energies, which are identified with 
the gravitational anomalous dimensions $\alpha_i$, once one
identifies time with the Liouville zero mode $\phi_0$~\cite{emn}.

The constants in $Q_0^2$ in the expression (\ref{ccd}) for $Q$ 
are fixed so as to provide the correct normalization 
for the kinetic energy of the recoiling $D$ particle
of the r.h.s. of (\ref{liouvener}). 
Such constants 
are attributed to properties of the equilibrium 
background space-time, in the limit $u^i\to 0$, 
and, as such, they are to be associated
with the specification of the rest energy of the background $D$-brane
foam. Equation (\ref{liouvener}), then,  
expresses the net change in energy
induced by the recoil kinetic energy of the $D$ brane affected by the 
highly-energetic particles.
We shall concentrate on this quantity from now on,
as a measure of the induced energy non-conservation in the 
observable particle subsystem.

Taking into account
spatial momentum conservation in the presence of the 
recoil $D$-particle deformation, which is rigorous in the 
context of the $\sigma$-model logarithmic conformal field theory
treatment of D-brane recoil~\cite{ms}, as discussed earlier, we may write:
\begin{equation} 
u=\sum _{i=1}^{N} { p_i \over M_D} 
\label{momconsdbrane}
\end{equation} 
from which we may write (\ref{liouvener}) in the form:
\begin{equation} 
\sum _{i=1}^{N}E_i= \frac{1}{2}\sum_{i=1}^{N} \frac{p_i^2}{M_D} 
+ \sum_{i<j} \frac{p_ip_j}{M_D} 
\label{lenc}
\end{equation}
We now consider an average $<< \dots >>$ over  
quantum (higher world-sheet genus) and statistical foam
effects. We concentrate first 
on the three-particle interaction in the presence of 
a recoiling $D$ particle, making the isotropy
assumption (\ref{isotropic}).
We then have from (\ref{lenc}): 
\begin{equation} 
\delta {\overline E}_D \equiv  
{\overline E}_1 - {\overline E}_2 - {\overline E_3} =
{M_D \over 2}<<u^2>>
\label{start}
\end{equation}
Using spatial momentum conservation (\ref{momconsn}), we then find
\begin{eqnarray}
\delta {\overline E}_D
&=& \frac{\xi_I}{M_D}\left( {\overline p}_2^2 +
{\overline p}_3^2 + {\overline p}_2 {\overline p}_3 \right) + \nonumber \\
&~& \frac{\xi_I}{M_D}<<{p}_2 {p}_3 >> -
\frac{\xi_I}{M_D}<<{p}_1 {p}_3>> - \nonumber \\
&~&\frac{\xi_I}{M_D}<<{p}_1 {p}_2>> + \frac{\xi_I}{M_D}\sum_{i=1}^{3}
(\Delta p_i)^2
\label{3decay}
\end{eqnarray}
where $\xi_I$ is a new numerical parameter ($\xi_I \ne \xi$ in
general) that characterizes the
effects of the statistical sum in $<< \dots >>$~\footnote{In the above, all the 
$E_i$ are positive, which explains the sign conventions.}.

To leading order in the inverse string or $D$-brane scale $M_D$,
the matrix element for any interaction agrees with 
that obtained in the framework of Special Relativistic 
Quantum Field Theory without gravitation, e.g.,
QED in the example of the $\gamma + \gamma \to e^+ + e^-$
process considered here. On the other hand, the 
back-reaction effects of the recoiling $D$ particle 
on the space-time are expected to modify the kinematics
of this field-theoretic result.
We examine their effects by factorizing multi-particle processes
as products of three-point vertices, and treating the latter
in a generic way without specifying 
the nature of the particles involved. We take into account 
spatial momentum conservation, but allow isotropic fluctuations, as
implied in our models of foam.

In the limit when $D$-particle recoil effects are ignored, 
one may
replace in (\ref{3decay}) 
products of the form $<<p_ip_j>>$ by the appropriate 
average (mean-field) momenta ${\overline p}_i{\overline p}_j$, up to 
quantum uncertainty effects of order 
$\chi_{ij} =<<p_ip_j>>-{\overline p}_i {\overline p}_j$:
\begin{equation} 
<<p_ip_j>> = {\overline p}_i{\overline p}_j + \chi_{ij}~,
\qquad \chi_{ij}= <<p_ip_j>>-({\overline p}_i {\overline p}_j)
\label{fluctuations}
\end{equation}
Then, using 
momentum conservation for the mean momenta:
${\overline p}_1 = {\overline p}_2 + {\overline p}_3$, it is
straightforward 
to see that the r.h.s. of (\ref{3decay}) consists only of 
the uncertainty variances $\chi_{ij}$, thereby implying 
the known flat space-time 
field-theory result 
of energy conservation during decay.

On the other hand, incorporating the recoil of the background  $D$
particle introduces a
distortion of space-time during the decay, 
changing the situation. 
In calculating such corrections 
in the presence of a random distribution 
of recoiling $D$ particles, we 
assume that the statistical average over the $D$-brane foam
generates a random distribution of angles
between the   
momentum of a particle produced during the decay, in our case particle
2, and the momenta of each of the external particles (1,3). This random
distribution
is compensated by the random distribution 
of angles of the recoiling $D$ particles, and hence spatial momentum 
is conserved on the average. Such processes provide minute 
fluctuating {\it
corrections}
to the mean-field relativistic field theory result 
on flat space-times.

In this case, the r.h.s. of the
pertinent three-particle interaction formula
(\ref{3decay}) becomes:
\begin{eqnarray} 
\delta {\overline E}_D = \frac{\xi_I}{M_D}{\overline p}_2^2 + 
\frac{\xi_I}{M_D}\sum_{i=1}^{3} (\Delta p_i)^2 +
\frac{\xi_I}{M_D}\chi_{13}
\label{3decayfinal}
\end{eqnarray}
where ${\overline p}_2$ is a typical momentum of a light particle
that may be exchanged in a four-body interaction,
$\chi_{13} \equiv <<p_1p_3>>-{\overline p}_1{\overline p}_3$,  
and again we took into account spatial 
momentum conservation (\ref{momconsn}). 
For all practical purposes, the 
contributions from the terms depending on 
variances $\Delta p_i$ and 
$\chi_{23}$ may be assumed negligible compared to 
$p_2^2/M_D$. In any case, they can only make {\it additive} contributions
to 
the r.h.s. of (\ref{3decayfinal}), and hence can never cancel
its first term.

Thus we arrive at the following conclusion: 

\noindent {\bf Theorem 2}:

{\it Observable momentum is conserved statistically and energy is violated
during particle interactions,
due to recoil effects in
$D$-particle space-time foam. The amount of energy violation is, 
in order of magnitude and up to quantum uncertainties, equal 
to the typical kinetic energy of the recoiling heavy (non-relativistic) 
$D$ particle, with energy being conserved in the complete system,
where the back-reaction on the space-time geometry of the 
recoiling $D$ particle is taken into account.}

\section{Implications for the GZK Cutoff}

As a non-trivial example that illustrates the physical
interest of the above analysis, consider now the 
process $\gamma_H \gamma_L \to e^+e^-$, where $\gamma_H$ is 
a highly energetic photon, and $\gamma_L$ an infrared background photon,
with energy $\omega \sim 0.025 eV$~\cite{protheroe}.
We recall that, in our framework, spatial momentum is conserved on the
average,
but the energy-momentum dispersion relations are modified
(\ref{averagedisp}). Also, as discussed in the previous section,
energy is violated in 3- and 4-particle decays and interactions 
(\ref{3decayfinal}).
We may, for our purposes, factorize the photon-photon scattering
four-point amplitude
into a product of two 3-body interactions mediated by the exchange 
of an off-shell electron with momentum $p_2$, as in standard QED.
The relevant energy-momentum (non-)conservation equations
are~\footnote{From now on, for simplicity we omit overlines in the
notation, since all the quantities below denote such averages.}:
\begin{eqnarray} 
E_1 + \omega  &=& E_2 + E_3 + \delta E_D~, \nonumber \\
p_1 - \omega &=& p_2 + p_3 
\label{conservequations}
\end{eqnarray}  
Inserting 
the modified energy-momentum dispersion relations for both massless and
massive particles in the first of the above equations,
and using the second, one obtains, in the case of highly-energetic
particles:
\begin{equation} 
2\omega + \frac{m_1^2}{2p_1} - 
\frac{m_2^2}{2p_2} - 
\frac{m_3^2}{2p_3} = \delta E_D^{(4)} + \frac{\xi}{M_D}
\left(p_2 p_3 + 
(p_2 + p_3 )\omega + \omega^2 \right)
\label{energynoncons}
\end{equation} 
For the purposes of the threshold calculation, we assume that $\omega <<
p_1\simeq E_{th} 
=2p_2=2p_3$, $m_1 \simeq 0$, $m_2=m_3=m_e$.
Then, from (\ref{energynoncons}) we obtain
\begin{equation} 
2\omega E_{th} - 2m_e^2 =E_{th}\left(\delta E_D^{(4)} + 
\frac{\xi}{M_D}(\frac{1}{2}E_{th}^2 + \omega)^2\right)
\label{threshold}
\end{equation}
where the $\omega$-dependent terms on the r.h.s of (\ref{threshold})
are negligible for our purposes.

The amount of energy violation $\delta E_D^{(4)}$ 
in a four-particle interaction such as $\gamma + \gamma \to e^+ + e^-$
or $N + \gamma \to \Delta \to N + \pi$ is
given by 
the sum of the corresponding violations in each of the two 
three-body interactions (\ref{3decayfinal}), assuming 
factorization of the relevant amplitudes, via virtual $e^\pm$ or $\Delta$
exchange, respectively. To leading order, and  
ignoring the stochastic quantum uncertainties in the 
prpagating particle energies, excitation of a
massive $D$ brane essentially determines $\delta E^{(4)}$:
\begin{equation} 
\delta E^{(4)} \simeq \frac{\xi_I}{2M_D}E_{th}^2 + \dots 
\label{ded4}
\end{equation}
where $\xi_I$ parametrizes energy loss during interactions, and is
{\it a priori} distinct from the propagation parameter $\xi$.

The fact that $\delta E^{(4)} \ge 0$, i.e., energy is {\it lost} during
particle decays in the presence of the $D$-particle foam, follows from the
sign of the flat-world-sheet anomalous dimension $-\epsilon^2/2$ of the
recoil deformation appearing inside the square root in the expression
(\ref{ad}) for $\alpha_D$. In our formalism, this sign is {\it positive},
because the corresponding operator is relevant in a world-sheet
renormalization-group sense~\cite{kmw}.  Should this operator have been
{\it irrelevant} instead, its anomalous dimension would have appeared with
the opposite sign inside (\ref{ad}), and one would have encountered a
(rather unphysical) situation in which there would be energy {\it gain} in
particle decays in the presence of $D$-particle foam. In such a case,
$\delta E^{(4)} <0$, and using the estimate (\ref{ded4}), which would
still be valid up to a change in sign, one might in principle have a
complete {\it cancellation} on the r.h.s. of the threshold equation
(\ref{threshold}), leaving intact the conventional Lorentz-invariant GZK
cutoff. However, our argument for the sign of $\xi_I$ excludes this
possibility.

In the high-energy limit $E_{th} >> \omega$,
(\ref{threshold}) becomes a cubic equation:
\begin{equation}
E_{th}-\frac{m_e^2}{\omega} \simeq \frac{(\xi_I +
\xi) E_{th}^3}{4 M_D\omega} + 
\dots  
\label{final}
\end{equation}
where the $\dots$ denote higher-order and
quantum-uncertainty effects. 
Equation (\ref{final}) always has a real solution
that determines the kinematic threshold.
For example, for the value $\omega \simeq 0.025$ eV, $m_e \sim 0.5$ MeV
and $M_D \sim 10^{19}$ GeV, two of the
three real solutions of the
cubic equation for $E_{th}$ are positive. 
The physical solution for $\xi = \xi_I = 1$ is $E_{th,2}/M_D \simeq 2.2
\times 10^{-15}$, which 
implies a non-trivial modification of the conventional  
GZK cutoff, namely an increase to
$\sim 20$ TeV, which is consistent with the observation of energetic
photons from Mk421. 

The fact that there is energy loss in particle interactions, as a result
of the
non-trivial recoil of excitations in the string/$D$-particle foam,
distinguishes our approach from
models~\cite{mestres,grillo,kifune,protheroe} where energy is assumed to
be conserved, as well as momentum. Our approach raises the GZK cutoff, but
it is lower than that predicted in models with $\xi_I = 0$.  Setting
$\delta E_D^{(4)}=0$ in (\ref{threshold}), the resulting cubic equation
for the threshold, analogous to (\ref{final}), would imply $E_{th} / M_D
\simeq 3.14 \times 10^{-15}$ if $\xi = 1$.

\section{Discussion}

In the equations (\ref{conservequations}) above, we have ignored quantum
uncertainties in the energies and momenta, assuming them to be small
compared to the mean-field effects of $D$-brane recoil. This is a feature
of the specific $D$-brane recoil model we have examined here~\cite{emn}.
In more generic approaches to quantum-gravitational foam, however, such an
assumption may not be valid, and quantum uncertainties associated in the
energies and momenta may be comparable to the terms leading to
modifications of the GZK cutoff discussed above. Even so, we
would not expect such stochastic uncertainties to dominate over the
modification of the usual dispersion relation.

However, such a
possibility has
recently been examined in~\cite{ng}, where energy and momentum
uncertainties of the following generic form were assumed: 
\begin{equation}
\Delta p_i \sim
\Delta E \sim \left(\frac{E}{M_P}\right)^\alpha E~, {\rm and} ~ \Delta p_i
- \Delta E =
\varepsilon \frac{E^{1 + \alpha}}{2 M_D^\alpha}
\label{ngequations}
\end{equation}
In the approach of~\cite{ng}, the parameters $\alpha$ and $\varepsilon$
are phenomenological, and should be bounded or determined by analyzing
high-energy cosmic-ray data. This case (for $\alpha = 1$) may be recovered
in our analysis above by setting $\xi \to 0$ in the appropriate
expressions (\ref{conservequations}, \ref{threshold}), etc., and keeping
only the terms associated with the quantum uncertainties, assuming the
energy-momentum dependent form (\ref{ngequations}) above. 

One way to distinguish the approach of~\cite{ng} from ours, in which
modified dispersion relations occur and probably dominate, is to look for
observable effects in the arrival times of energetic photons from distant
astrophysical sources such as GRBs~\cite{aemn,nature}. In our approach, in
which stochastic fluctuations are not expected to dominate, one would
expect searches for different arrival times for photons in {\it different}
energy ranges to be fruitful, for example pulses emitted by GRBs in
different energy channels~\cite{nature,mitsou}. On the other hand, in
approaches to quantum-gravitational foam like that of~\cite{ng}, in which
quantum uncertainties in energies and momenta are dominant, one should
observe energy-dependent spreads in the arrival times of photons {\it of
the same energy}, which could be probed by comparing the widths of pulses
emitted by GRBs in different energy channels~\cite{emnstoch,mitsou}.

A final remark is that our string approach to quantum space-time foam is
compatible with the general principles of relativity, despite the
appearance of a fundamental (Planckian) mass scale $M_D$.  In string
theory, this mass scale is an observer-independent quantity {\it by
construction}. Our modifications of Lorentz symmetry, as expressed in the
modified dispersion relations discussed above, pertain to a specific
background of string theory, so such Lorentz violations should be
considered as {\it spontaneous}.  In the Liouville approach to
time~\cite{emn}, although the time coordinate is special, in the sense of
being associated with a renormalization-group scale on the world sheet of
the string, and thus irreversible~\cite{zam}, there is no contradiction
with general coordinate invariance, given that Liouville strings do embody
that principle~\cite{emn} in their $\sigma$-model framework. The
irreversibility of the Liouville time is no stranger than time
irreversibility in cosmological models, which are compatible with general
coordinate invariance, being specific solutions of the gravitational field
equations. 

In this respect, we question the need for extra axioms~\cite{amelino2000}
in order to formulate theories with an observer-independent minimal
length.  In our stringy approach, where the induced violations of Lorentz
symmetry are due to the (dynamical) choice of background, there is a
frame-dependence of the energy scale associated with modifications of the
light velocity, but this is a reflection of the spontaneous violation of
Lorentz invariance, and the underlying (Planckian) mass scale is
universal.  The ensuing predictions are compatible with the general
principles of relativity. 

\section*{Acknowledgements}

The work of D.V.N. is partially supported by DOE grant
DE-F-G03-95-ER-40917.  N.E.M. and D.V.N. also thank H. Hofer for his
interest and support.

\end{document}